# Title: Oxygen isotopic evidence for vigorous mixing during the Moon-forming Giant Impact

**Authors:** Edward D. Young[1]*, Issaku E. Kohl[1]*, Paul H. Warren[1], David C. Rubie[2], Seth A. Jacobson[2,3], and Alessandro Morbidelli[3].

**Affiliations:**

[1]Department of Earth, Planetary, and Space Sciences, University of California Los Angeles, Los Angeles, CA, USA.
[2]Bayerisches Geoinstitut, University of Bayreuth, D-95490 Bayreuth, Germany.
[3]Laboratoire Lagrange, Universite' de Nice - Sophia Antipolis, Observatoire de la Cote d'Azur, CNRS, 06304 Nice, France.

*Correspondence to: eyoung@epss.ucla.edu, ikohl@epss.ucla.edu

**Abstract**: Earth and Moon are shown here to be composed of oxygen isotope reservoirs that are indistinguishable, with a difference in $\Delta'^{17}O$ of $-1$ +/− 5ppm (2se). Based on these data and our new planet formation simulations that include a realistic model for oxygen isotopic reservoirs, our results favor vigorous mixing during the giant impact and therefore a high-energy high-angular-momentum impact. The results indicate that the late veneer impactors had an average $\Delta'^{17}O$ within approximately 1 ‰ of the terrestrial value, suggesting that these impactors were water rich.

**One Sentence Summary:** Moon and Earth derived from indistinguishable oxygen isotopic reservoirs, placing constraints on the relative fractions of Moon-forming impactor in Earth and Moon.

**Main Text:** The Moon is thought to be the consequence of a giant collision between the proto-Earth and a planetary embryo or proto-planet (named Theia, "mother of the Moon") ~ $10^7$ to $10^8$ years after the birth of the solar system (*1, 2*). However, the unique oxygen isotopic signatures of Solar System bodies (*3, 4*) has presented a problem for the impact hypothesis for the formation of the Moon (*5, 6*). In order to create an iron-poor Moon and simultaneously reproduce the angular momentum of the Earth-Moon system, early models required a glancing blow by a Mars-sized impactor that resulted in Moon being composed mainly of impactor material (*7*). Therefore, in the general case, Moon and Earth should not be identical in their oxygen isotopic compositions. Nonetheless, until recently, Moon and Earth have been found to be indistinguishable in their oxygen isotope ratios (*8-10*). In later models, a larger Theia/proto-Earth size ratio leads to more thorough mixing of the silicate portions of the target and impactor, offering a solution to this conundrum (*11*), although at the expense of the need to shed significant angular momentum from the system via orbital resonances (*12*).

Oxygen reservoirs comprising rocky bodies of the Solar System are characterized by unique relative concentrations of oxygen isotopes. These relative concentrations are customarily represented by $\Delta^{17}O$, the departure in $^{17}O/^{16}O$ relative to a given $^{18}O/^{16}O$ under the assumption that these two isotope ratios covary as a consequence of mass-dependent isotope fractionation. Differences in oxygen isotope ratios in geological materials are small and are therefore reported as fractional differences. It is common to replace the per mil (‰) fractional differences with $\delta'^{17}O = 10^3 \ln(^{17}R/^{17}R_o)$ and $\delta'^{18}O = 10^3 \ln(^{18}R/^{18}R_o)$ (*13*) values where $^{17}R$ (or $^{18}R$) is $^{17}O/^{16}O$ (or $^{18}O/^{16}O$) and the subscript o refers to the initial isotope ratios (e.g., those characterizing bulk Earth) (*14*). These latter values are numerically nearly equivalent to the fractional differences but have the useful property that they are related linearly by the mass fractionation exponent $\beta$. The fractionation exponent $\beta$ has a value near ½ determined by the masses of the three isotopes of oxygen. The exact values for $\beta$ depend on the processes involved in fractionation (*13*). With these definitions $\Delta^{17}O$ is written as

$$\Delta'^{17}O = \delta'^{17}O - \beta\, \delta'^{18}O \ . \qquad (1)$$

A positive $\Delta'^{17}O$ signifies that a reservoir is enriched in $^{17}O$ relative to Earth while a negative value signifies that a reservoir is relatively depleted in $^{17}O$ compared with expectations from mass fractionation. At the scale of individual mineral grains, Solar System materials exhibit variations in $\Delta'^{17}O$ spanning ~ 200 ‰ (*15*). The dispersion in $\Delta'^{17}O$ decreases drastically with mass. Differences in $\Delta'^{17}O$ among meteorite whole-rock samples are about 5 to 8 ‰ (*4, 16*), representing parent asteroids with masses of ~ $10^{15}$ to $10^{17}$ kg. Differences between differentiated bodies with metal cores and silicate mantles are smaller still: Mars ($6.4 \times 10^{23}$ kg) has a $\Delta'^{17}O$ value of about +0.3 ‰ while Vesta ($2.6 \times 10^{20}$ kg) has a value of −0.25 ‰ (*17, 18*). The reduced dispersion in $\Delta'^{17}O$ with mass evidently reflects averaging as smaller rocky bodies coalesced to form larger bodies in the Solar System (*19, 20*). Historically, the identical $\Delta'^{17}O$ values for Earth and Moon have stood out against this backdrop of variability in the Solar System.

Most recently, it was suggested that Moon has a greater $\Delta'^{17}O$ than Earth by 12 +/– 3 ppm *(21)*. The significance of this finding can be gauged with reference to Fig. 1A in which contours for $\Delta'^{17}O_{Moon} - \Delta'^{17}O_{Earth}$ are plotted as functions of the difference in $\Delta'^{17}O$ between Theia and the proto-Earth and the difference in the fractions of Moon and Earth inherited from Theia. The mass-balance equation plotted is

$$x_{Theia, Moon} - x_{Theia, Earth} = \frac{\Delta'^{17}O_{Moon} - \Delta'^{17}O_{Earth}}{\Delta'^{17}O_{Theia} - \Delta'^{17}O_{proto-Earth}} \quad (2)$$

where $x_{Theia,i}$ refers to the oxygen fraction of body *i* derived from Theia (essentially mass fractions of the bulk silicate portions of the bodies). For convenience we also use the fractional difference $\delta_{Theia}$ rather than the absolute difference in Equation (2), so that

$$\delta_{Theia} = (x_{Theia, Moon} - x_{Theia, Earth}) / x_{Theia, Earth} . \quad (3)$$

The significance of a difference in oxygen isotopic composition between Moon and Earth depends on the fraction of Theia contained within Earth (Equations 2 and 3). Four recent proposed giant impact scenarios (*5, 11, 12, 22*) predict disparate differences in the Theia fractions in Moon and Earth (Fig. 1A). If the difference in $\Delta'^{17}O$ between Theia and the proto-Earth was zero, there is no oxygen isotope constraint on $\delta_{Theia}$ (Fig. 1A). Similarly, if Earth and Moon are composed of precisely the same concentrations of Theia, there is no constraint on differences in $\Delta'^{17}O$ between Theia and the proto-Earth. Contours for $\Delta'^{17}O_{Moon} - \Delta'^{17}O_{Earth}$ in Fig. 1A show all cases in between.

The positive $\Delta'^{17}O$ of 12 +/– 3 ppm for the Moon *(21)* *requires* a difference in the proportions of Moon and Earth composed of remnants of Theia. This is clear because the contours representing this range of values (violet regions) do not include the center of the diagram. For a Mars-sized differentiated body with $\Delta'^{17}O \sim$ +/– 0.3 ‰ (e.g., Mars and Vesta), the difference in Theia contents between Moon and Earth is +/– 50% or more (Fig. 1A). For the case of a proto-Earth sized Theia, the result is a difference of +/– 8% or more (Fig. 1A). Alternatively, assuming enstatite-chondrite like material better represents the terrestrial planet-forming region (*23, 24*), differences in oxygen isotope ratios between Theia and proto-Earth would have been smaller (~0.1 ‰ (*21, 25*)), and the lunar $\Delta'^{17}O$ of 12 +/– 3 ppm (*21*) requires $\delta_{Theia}$ values of 150% and 30% for the Mars and proto-Earth-sized impactors, respectively (Fig. 1A). Such large $\delta_{Theia}$ values would effectively remove the constraint imposed by oxygen isotopes that the Earth-Moon system was well-mixed.

We analyzed seven Apollo 12, 15, and 17 lunar samples and one lunar meteorite and compared their $^{17}O/^{16}O$ and $^{18}O/^{16}O$ isotope ratios with those for a suite of terrestrial igneous samples. The 1 to 4 milligram lunar samples include high-Ti mare basalts, low-Ti Mg-rich olivine cumulate basalts, a quartz normative basalt, and a highland anorthositic troctolite (Table S1). The terrestrial samples include San Carlos mantle xenolith olivines, San Carlos mantle xenolith spinels, Mauna Loa basalt samples, Mauna Loa olivine separates, an anorthosite from the Bushveld complex, and a sample of Gore Mountain metamorphic garnet. Our analyses (Table 1) were obtained using infrared laser heating (*26*) modified to include $F_2$ as the fluorinating agent and purification of the analyte $O_2$ gas for analysis of both $^{17}O/^{16}O$ and $^{18}O/^{16}O$ (*27*). We have improved our precision compared with many previous efforts by more thoroughly desiccating samples prior to analysis and by regular re-balancing of standard and sample ion beam intensities throughout the mass spectrometer analyses (*28*). We analyzed a range of lunar and terrestrial sample lithologies to account for the fact that $\beta$ values vary with process (*13, 29, 30*). We use the traditional Standard Mean Ocean Water (SMOW) as the reference for $\delta'^{18}O$, but we use San Carlos (SC) olivine as the reference for $\Delta'^{17}O$ when characterizing oxygen isotope reservoirs of rocks (*28*). We adopt a typical igneous $\beta$ of 0.528 passing through the mean value for San Carlos olivine as our reference fractionation line for calculating $\Delta'^{17}O$ (*28*).

Lunar basalts are relatively high in $\delta'^{18}O$ compared with SC olivine and terrestrial basalts (Fig. 2). Nonetheless, the basalts show no clear deviation from the reference $\beta$ of 0.528, allowing direct comparison of $\Delta'^{17}O$ values for these materials. There is no discernible difference between lunar basalts and SC olivine, with the former yielding a $\Delta'^{17}O$ of –0.001 +/– 0.002 (1se) for powders and 0.000 +/– 0.003 (1se) for fused beads. The mean for all mafic terrestrial samples, representing terrestrial mantle and its melt products, is 0.000 +/- 0.001 ‰ (1se). Adding in quadrature the analytical uncertainty in the SC olivine and the standard error for the lunar samples yields a difference between lunar basalt and SC olivine of –0.001 +/– 0.0048 ‰ (−1 ppm +/− 4.8 ppm, 2se), indistinguishable from zero. Other mafic terrestrial whole rocks and olivines are within this uncertainty range (Table 1), and we conclude that there is no resolvable difference in $\Delta'^{17}O$ between lunar mantle melts represented by these basalts and terrestrial mantle and melts.

Our result does not agree with the conclusions of Herwartz et al. (*21*). Measurements on the one sample common to both studies (12018) agree within uncertainties when compared in the same reference frame (*(28)*, Fig. S2). It is therefore possible that the vicissitudes of sample selection are a plausible explanation for the difference between the studies.

The lunar highland sample has a statistically significantly lower $\Delta'^{17}O$ value of − 0.016 +/− 0.003 (1se) ‰ (or −16 +/− 3 ppm ), similar to a previous study (*8*). However, the terrestrial anorthosite sample has a similarly low value (Table 1). The low $\Delta'^{17}O$ values for both the terrestrial and lunar highland anorthosites (anorthostic troctolite) imply a mass fractionation process related to formation of this rock type that results in low $\Delta'^{17}O$ values (Fig. 2). The low $\Delta'^{17}O$ value for the lunar highland rocks is not evidence for a distinction between the oxygen pools for the Moon and Earth because these samples are in the mass-fractionation envelope for Earth (Fig. 2) and low $\Delta'^{17}O$ values are found in both terrestrial and lunar anorthosite-like rocks. One terrestrial mantle spinel sample also shows a measurable deviation from the $\beta = 0.528$ reference, implying a relatively low $\beta$ value (Fig. 2).

Of course in all cases, invoking no difference in oxygen isotope ratios between Theia and proto-Earth results in no constraints on the relative Theia concentrations in Moon and Earth. Is this a viable scenario? We can assess the purely statistical feasibility of two proto-planetary bodies having identical oxygen isotope ratios using the central limit theorem (*19*). Results suggest that a purely random sampling of asteroid-like materials would lead to variations in $\Delta'^{17}O$ among planetesimals of ~ 3 ppm (*28)*. However, the larger difference between Earth and Mars testifies to the fact that $\Delta'^{17}O$ was not distributed randomly in small bodies across the inner Solar System.

Differences in $\Delta'^{17}O$ values between Theia and the proto-Earth have expected values of 0.15‰ (*31*) or 0.05 ‰ (*32*) based on two recent N-body simulations of standard terrestrial planet formation scenarios with hypothesized gradients in $\Delta'^{17}O$ across the inner Solar System. We used a planetary accretion model (*33*) that utilizes N-body accretion simulations based on the Grand Tack scenario (*34*). Our model differs from previous efforts in that we strictly limit our analysis to simulations that closely reproduce the current masses and locations of Earth and Mars and the oxidation state of Earth's mantle, we use a multi-reservoir model to describe the initial heliocentric distribution of oxygen isotopes, and we include the effects of mass accretion subsequent to the Moon-forming impact (*28*). An example simulation (Fig. 3), and others like it, shows that the $\Delta'^{17}O$ values of the colliding bodies rise together as the average $\Delta'^{17}O$ values increase during accretion. Incorporation of more material from greater distances from the Sun as accretion proceeds accounts for the rise. Large planets like Earth and Venus reflect an average of many embryos and planetesimals and so exhibit similar $\Delta'^{17}O$ values with time, while stranded embryos averaging fewer components, like Mars, show greater variation.

The cumulative distribution of $\Delta'^{17}O$ differences between Theia and proto-Earth is shown for 236 simulations of planet growth (*35*) in Fig. 1B. The median $\Delta'^{17}O_{Theia} - \Delta'^{17}O_{proto\text{-}Earth}$ is nearly 0 in these calculations for all simulations (Fig. 1B). However, our median predicted $\Delta'^{17}O_{Theia} - \Delta'^{17}O_{proto\text{-}Earth}$ is +0.1 ‰ if we restrict our analysis to those simulations consistent with adding ≤ 1% by mass of a "late veneer" of primitive material post Moon-forming impact as required by geochemical constraints. This median value combined with our measurement of $\Delta'^{17}O_{Moon} - \Delta'^{17}O_{Earth}$ corresponds to $\delta_{Theia}$ of +20% to –60 % for the Mars-sized impactor scenario and +8% to –12 % in the proto-Earth sized impactor scenarios. The corresponding values for $\delta_{Theia}$ using the previous 12 +/– 3 ppm difference between Moon and Earth $\Delta'^{17}O$ values (*21*) are +80 to +180% and +16 to +36 %, respectively (Fig. 1A). The new measurements presented here are consistent with Earth and the Moon having near-identical Theia contents.

Indistinguishable $\Delta'^{17}O$ values of Moon and Earth to the 5 ppm level of uncertainty is consistent with a Moon-forming impact that thoroughly mixed and homogenized the oxygen isotopes of Theia and proto-Earth, a conclusion not permitted by the previous estimate of a measurable difference in oxygen isotopic composition between these bodies.

Our interpretation is bolstered by the amount and composition of the late veneer of primitive bodies required by the excess of highly siderophile elements (HSEs) in the silicate portion of Earth (*36*). A disproportionately larger flux of late-veneer planetesimals is implied by a higher average $^{182}W/^{184}W$ for Moon than for Earth and by previous estimates for the apparent differences in HSE concentrations between the terrestrial and lunar mantles (*37*). The interpretation of these data is that Moon and Earth began with the same W isotopic ratios but that the Earth inherited a greater fraction of low $^{182}$W/$^{184}$W material in the form of chondritic planetesimals after the Moon-forming impact (*38, 39*). If we adopt the conclusion from the W isotopes that the Earth-Moon system was well mixed as a result of the Moon-forming impact, then the nearly identical $\Delta'^{17}$O values of Moon and Earth constrain the identity of the late-veneer impactors by their oxygen isotope ratios.

Estimates for the Earth/Moon ratio of the late-veneer mass fluxes range from ~ 200 to 1200 (*37, 40, 41*). Using a late-veneer flux to Earth of $2\times10^{22}$ kg indicated by the concentrations of HSEs in Earth's mantle (*37*) and a conservative maximum Earth/Moon flux ratio of 1200 (*41*), the difference in late-veneer fractions comprising the silicate Earth and Moon is 0.00447. This number can be converted to the mean oxygen isotopic composition of late-veneer impactors using our measured value for $\Delta'^{17}$O$_{Moon}$ – $\Delta'^{17}$O$_{Earth}$ of effectively zero (*29*). In this case the LV impactors would be similar to enstatite chondrites in having $\Delta'^{17}$O values within ~ 0.2 ‰ or less of Earth (*25*). It is worth noting that enstatite chondrites have some other isotopic affinities to Earth (e.g., *37*). Alternatively, with our maximum permitted $\Delta'^{17}$O$_{Moon}$ – $\Delta'^{17}$O$_{Earth}$ of ± ~5 ppm, the calculated $\Delta'^{17}$O value for the LV is ±1.1 ‰. This value encompasses aqueously altered carbonaceous chondrites, some ordinary chondrites, and mixtures thereof. For comparison, the same calculation using the 12 ppm difference between Moon and Earth yields an LV $\Delta'^{17}$O of −2.7 ‰, suggesting that the impactors were composed mainly of relatively unaltered and dry carbonaceous chondrites (*4*). Our result suggests that if the late veneer was composed mainly of carbonaceous chondrites, the parent bodies must have included significant fractions of high-$\Delta'^{17}$O water either in the form of aqueous alteration minerals or as water ice.

The timing of the late veneer also supports our interpretation that our results indicate thorough mixing during the giant impact rather than identical $\Delta'^{17}$O values for Theia and proto-Earth. The HSE restriction on the maximum mass of the late veneer requires a late Moon-forming giant impact. Because planets grow by accreting nearest neighbors first, a later impact suggests that the candidate giant impactors were unlikely to have formed adjacent the proto-Earth. Theia and the proto-Earth are less likely to have had identical $\Delta'^{17}$O in this circumstance because dynamical stirring brings together planetesimals and embryos with different oxygen reservoirs with time.

**Acknowledgments:** We are grateful to NASA Johnson Space Center for approving use of the Apollo samples for this study. EDY acknowledges support from a grant from the NASA

Emerging Worlds program (NNX15AH43G). DCR, SAJ, and AM acknowledge support from the European Research Council Advanced Grant "ACCRETE" (contract number 290568). The full data table for this study can be found in the Online Material.

**Supplementary Materials:**

Supplementary Text

Figure S1, S2, S3, S4, S5, S6, and S7

Tables S1 and S2

**Table 1**. Summary of oxygen isotope data for lunar and terrestrial samples. Delta values are in logarithmic form as defined in the text.

| Sample | $\delta^{17}O'$ | 1 se | $\delta^{18}O'$ | 1 se | $\Delta^{17}O'$ | 1 se |
|---|---|---|---|---|---|---|
| **Lunar basalt** | | | | | | |
| Avg (N=8) | 3.004 | | 5.691 | | -0.001 | |
| Std deviation | 0.090 | | 0.172 | | 0.005 | |
| Std error | 0.032 | | 0.061 | | 0.002 | |
| **Lunar basalt -fused beads** | | | | | | |
| Avg (N=4) | 2.940 | | 5.572 | | 0.000 | |
| Std deviation | 0.133 | | 0.245 | | 0.006 | |
| Std error | 0.067 | | 0.123 | | 0.003 | |
| **Lunar troctolite** | | | | | | |
| Avg (N=2) | 3.178 | | 6.050 | | -0.016 | |
| Std deviation | 0.009 | | 0.026 | | 0.005 | |
| Std error | 0.007 | | 0.019 | | 0.003 | |
| **San Carlos olivine** | | | | | | |
| Avg (N=17) | 2.711 | | 5.134 | | 0.000 | |
| Std deviation | 0.072 | | 0.134 | | 0.005 | |
| Std error | 0.017 | | 0.033 | | 0.001 | |
| **Mauna Loa olivine** | | | | | | |
| Avg (N=4) | 2.736 | | 5.189 | | -0.004 | |
| Std deviation | 0.090 | | 0.170 | | 0.001 | |
| Std error | 0.045 | | 0.085 | | 0.001 | |
| **Mauna Loa whole-rock samples** | | | | | | |
| Avg (N=5) | 2.796 | | 5.298 | | -0.002 | |
| Std deviation | 0.031 | | 0.063 | | 0.003 | |
| Std error | 0.014 | | 0.028 | | 0.001 | |
| **San Carlos spinel** | | | | | | |
| Avg (N=2) | 2.171 | | 4.104 | | 0.004 | |
| Std deviation | 0.135 | | 0.285 | | 0.015 | |
| Std error | 0.096 | | 0.202 | | 0.011 | |

**Bushveld anorthosite**

| | | | |
|---|---|---|---|
| Avg (N=2) | 3.522 | 6.694 | -0.012 |
| Std deviation | 0.002 | 0.002 | 0.001 |
| Std error | 0.001 | 0.002 | 0.000 |

**Gore Mountain garnet**

| | | | |
|---|---|---|---|
| Avg (N=2) | 3.174 | 6.020 | -0.004 |
| Std deviation | 0.017 | 0.026 | 0.003 |
| Std error | 0.012 | 0.019 | 0.002 |

**San Carlos olivine - doubly-distilled $O_2$**

| | | | |
|---|---|---|---|
| Avg (N=4) | 2.735 | 5.185 | -0.002 |
| Std deviation | 0.072 | 0.141 | 0.002 |
| Std error | 0.042 | 0.082 | 0.001 |

1. W. K. Hartmann, D. R. Davis, Satellite-sized planetesimals and lunar origin. *Icarus* **24**, 504-515 (1975).
2. T. Kleine *et al.*, Hf-W chronology of the accretion and early evolution of asteroids and terrestrial planets. *Geochemica et Cosmochimica Acta* **73**, 5150-5188 (2009).
3. R. N. Clayton, Oxygen isotopes in meteorites. *Annual Reviews in Earth and Planetary Science* **21**, 115-149 (1993).
4. R. N. Clayton, N. Onuma, T. K. Mayeda, A classification of meteorites based on oxygen isotopes. *Earth and Planetary Science Letters* **30**, 10-18 (1976).
5. R. M. Canup, Lunar-forming collisions with pre-impact rotation. *Icarus* **196**, 518-538 (2008).
6. K. Pahlevan, D. J. Stevenson, Equilibration in the aftermath of the lunar-forming giant impact. *Earth and Planetary Science Letters* **262**, 438-449 (2005).
7. R. M. Canup, E. Asphaug, Origin of the Moon in a giant impact near the end of the Earth's formation. *Nature* **412**, 708-712 (2001).
8. U. H. Wiechert *et al.*, Oxygen isotopes and the Moon-forming giant impact. *Science* **294**, 345-348 (2001).
9. L. J. Hallis *et al.*, The oxygen isotope composition, petrology and geochemistry of mare basalts: Evidence for large-scale compositional variation in the lunar mantle. *Geochemica et Cosmochimica Acta* **74**, 6885-6899 (2010).
10. M. J. Spicuzza, J. M. D. Day, L. A. Taylor, J. W. Valley, Oxygen isotope constraints on the origin and differentiation of the Moon. *Earth and Planetary Science Letters* **253**, 254-265 (2007).
11. R. M. Canup, Forming a Moon with an Earth-like composition via a giant impact. *Science* **338**, 1052-1055 (2012).
12. M. Cuk, S. T. Stewart, Making the Moon from a fast-spinning Earth: a giant impact followed by resonant despinning. *Science* **338**, 1047-1052 (2012).
13. E. D. Young, A. Galy, H. Nagahara, Kinetic and equilibrium mass-dependent isotope fractionation laws in nature and their geochemical and cosmochemical significance. *Geochimica et Cosmochimica Acta* **66**, 1095-1104 (2002).
14. C. R. McKinney, J. M. McCrea, S. Epstein, H. A. Allen, H. C. Urey, Improvements in mass spectrometers for the measurement of small differences in isotope abundance ratios. *The Review of Scientific Instruments* **21**, 724-730 (1950).
15. N. Sakamoto *et al.*, Remnants of early Solar System water enriched in heavy oxygen isotopes. *Science* **317**, 231-233 (2007).
16. M. K. Weisberg *et al.*, The Carlisle Lakes-type chondrites: a new grouplet with high delta-17O and evidence for nebular oxidation. *Geochemica et Cosmochimica Acta* **55**, 2657-2669 (1991).
17. I. A. Franchi, I. P. Wright, A. S. Sexton, C. T. Pillinger, The oxygen-isotopic composition of Earth and Mars. *Meteoritics & Planetary Science* **34**, 657-661 (1999).
18. R. N. Clayton, T. K. Mayeda, Oxygen isotope studies in achondrites. *Geochimica et Cosmochimica Acta* **60**, 1999-2017 (1996).
19. J. A. I. Nuth, H. G. M. Hill, Planetary accretion, oxygen isotopes, and the central limit theorem. *Meteoritics & Planetary Science* **39**, 1957-1965 (2004).
20. V. S. Safronov, Sizes of the largest bodies falling onto the planets during their formation. *Soviet Astronomy* **9**, 987-991 (1966).

21. D. Herwartz, A. Pack, B. Friedrichs, A. Bischoff, Identification of the giant impactor Theia in lunar rocks. *Science* **344**, 1146-1150 (2014).
22. A. Reufer, M. M. M. Meier, W. Benz, R. Wieler, A hit-and-run giant impact scenario. *Icarus* **221**, 296-299 (2012).
23. S. B. Jacobsen, M. I. Petaev, S. Huang, paper presented at the American Geophysical Union Fall Meeting, San Francisco, CA, 2012.
24. L. R. Nittler *et al.*, The major-element composition of Mercury's surface from MESSENGER X-ray spectrometry. *Science* **333**, 1847-1850 (2011).
25. J. Newton, I. A. Franchi, C. T. Pillinger, The oxygen-isotopic record in enstatite meteorites. *Meteoritical & Planetary Science* **35**, 689-698 (2000).
26. Z. D. Sharp, A laser-based microanalytical method for the in situ determination of oxygen isotope ratios of silicates and oxides. *Geochemica et Cosmochimica Acta* **54**, 1353-1357 (1990).
27. E. D. Young, H. Nagahara, B. O. Mysen, D. M. Audet, Non-Rayleigh oxygen isotope fractionation by mineral evaporation: Theory and experiments in the system $SiO_2$. *Geochimica et Cosmochimica Acta* **62**, 3109-3116 (1998).
28. Materials and methods are available as supplementary materials on Science Online
29. X. Cao, Y. Liu, Equilibrium mass-dependent fractionation relationships for triple oxygen isotopes. *Geochemica et Cosmochimica Acta* **75**, 7435-7445 (2011).
30. J. Matsuhisa, J. R. Goldsmith, R. N. Clayton, Mechanisms of hydrothermal crystallisation of quartz at 250 C and 15 kbar. *Geochimica et Cosmochimica Acta* **42**, 173-182 (1978).
31. N. A. Kaib, N. B. Cowan, The feeding zones of terrestrial planets and insights into Moon formation. *Icarus* **252**, 161-174 (2015).
32. A. Mastrobuono-Battisti, H. B. Perets, S. N. Raymond, A primordial origin for the compositional similarity between Earth and the Moon. *Nature* **520**, 212-215 (2015).
33. D. C. Rubie *et al.*, Accretion and differentiation of the terrestrial planets with implications for the compositions of early-formed Solar System bodies and accretion of water. *Icarus* **248**, 89-108 (2015).
34. K. J. Walsh, A. Morbidelli, S. N. Raymond, D. P. O'Brien, A. M. Mandell, A low mass for Mars from Jupiter's early gas-driven migration. *Nature* **475**, 206-209 (2011).
35. S. A. Jacobson, A. Morbidelli, Lunar and terrestrial planet formation in the Grand Tack scenario. *Royal Society of London Philosophical Transactions Series A* **372**, 0174 (2014).
36. C.-L. Chou, in *Proceedings of the Lunar and Plantary Science Conference*. (Houston, TX, 1978), vol. 9, pp. 219-230.
37. R. J. Walker *et al.*, In search of late-stage planetary building blocks. *Chemical Geology* **411**, 125-142 (2015).
38. T. S. Kruijer, T. Kleine, M. Fischer-Godde, P. Spring, Lunar tungsten isotopic evidence for the late veneer. *Nature* **520**, 534-537 (2015).
39. M. Touboul, I. S. Puchtel, R. J. Walker, Tungsten isotopic evience for disproprtional late accretion to the Earth and Moon. *Nature* **520**, 530-533 (2015).
40. H. E. Schlichting, P. H. Warren, Q.-Z. Yin, The last stages of terrestrial planet formation: dynamical friction and the late veneer. *The Astrophysical Journal* **752:8**, 8 pp. (2012).
41. W. F. Bottke, R. J. Walker, J. M. D. Day, D. Nesvorny, L. Elkins-Tanton, Stochastic late accretion to Earth, the Moon, and Mars. *Science* **330**, 1527-1530 (2010).

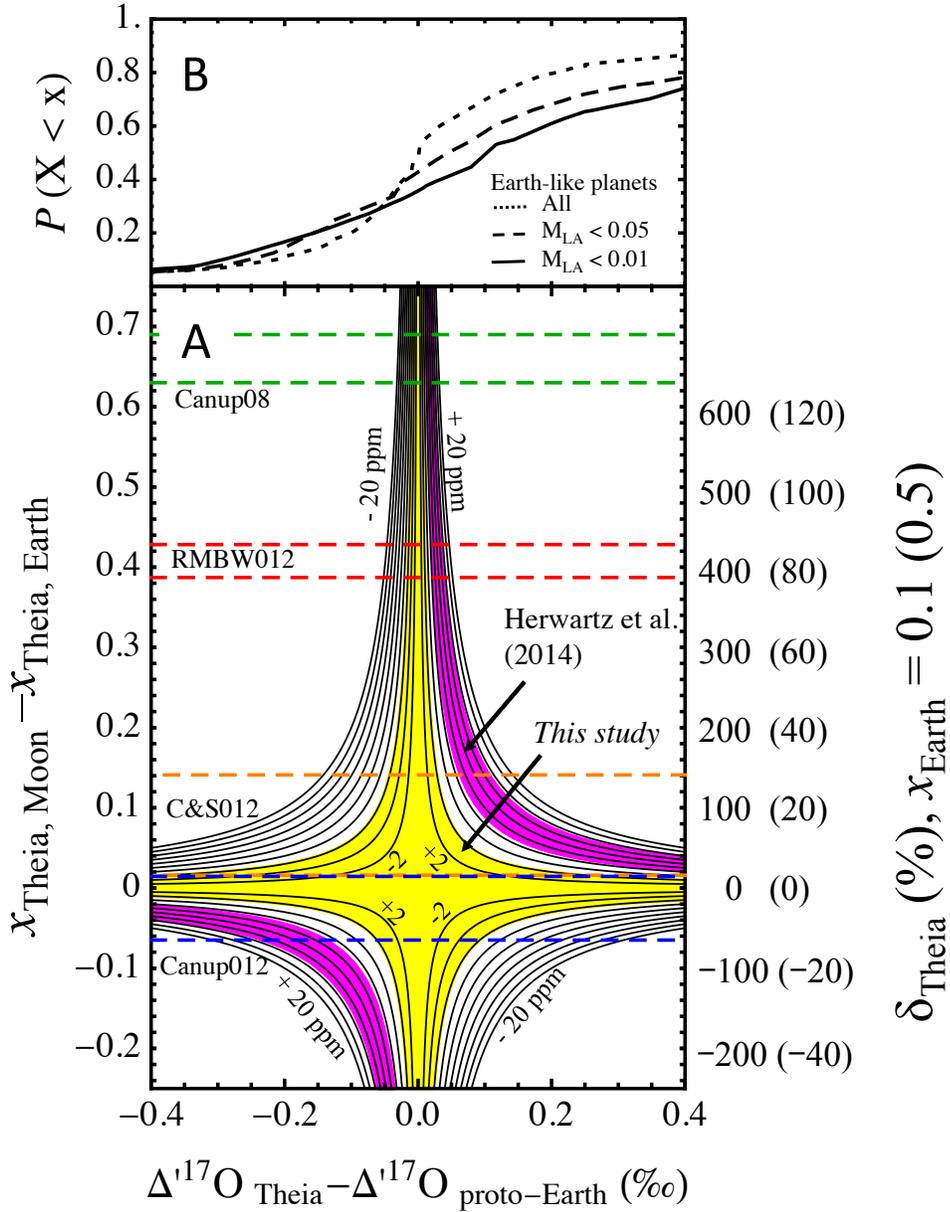

**Figure 1.** Oxygen isotope mass balance diagram. (A) Contours of $\Delta'^{17}O_{Moon} - \Delta'^{17}O_{Earth}$ in ppm versus fractional differences in Theia content of the bulk silicate Moon and Earth and $\Delta'^{17}O_{Theia} - \Delta'^{17}O_{proto-Earth}$. The contour interval is 2 ppm. The violet region indicates the contour intervals consistent with the $\Delta'^{17}O_{Moon} - \Delta'^{17}O_{Earth}$ reported by Herwartz et al. (2014). The yellow region encompasses the contours consistent with our data +/-2 standard error. Corresponding values for $\delta_{Theia}$ are shown at right. One set of $\delta_{Theia}$ values applies if the fraction of the present-day bulk silicate Earth composed of Theia is 0.1, while the values in parentheses apply where the fraction of Theia in present-day Earth is 0.5. For comparison, the ranges in Theia contents of the Moon and Earth for four published simulated moon-forming impact scenarios are shown as dashed horizontal lines. The models include the "canonical" model requiring no subsequent angular momentum loss by Canup, (2008, Canup08), the hit-and-run model of Reufer et al., (2012, RMBW012), and the high angular momentum scenarios, including Cuk & Stewart (2012, C&S012) and Canup (2012, Canup012). (B) The cumulative probability for $\Delta'^{17}O_{Theia} - \Delta'^{17}O_{proto-Earth}$ in per mil based on simulations in this study. Three cases are shown: those with late accreted mass to Earth < 5%, those with late accreted mass < 1%, and all simulations.

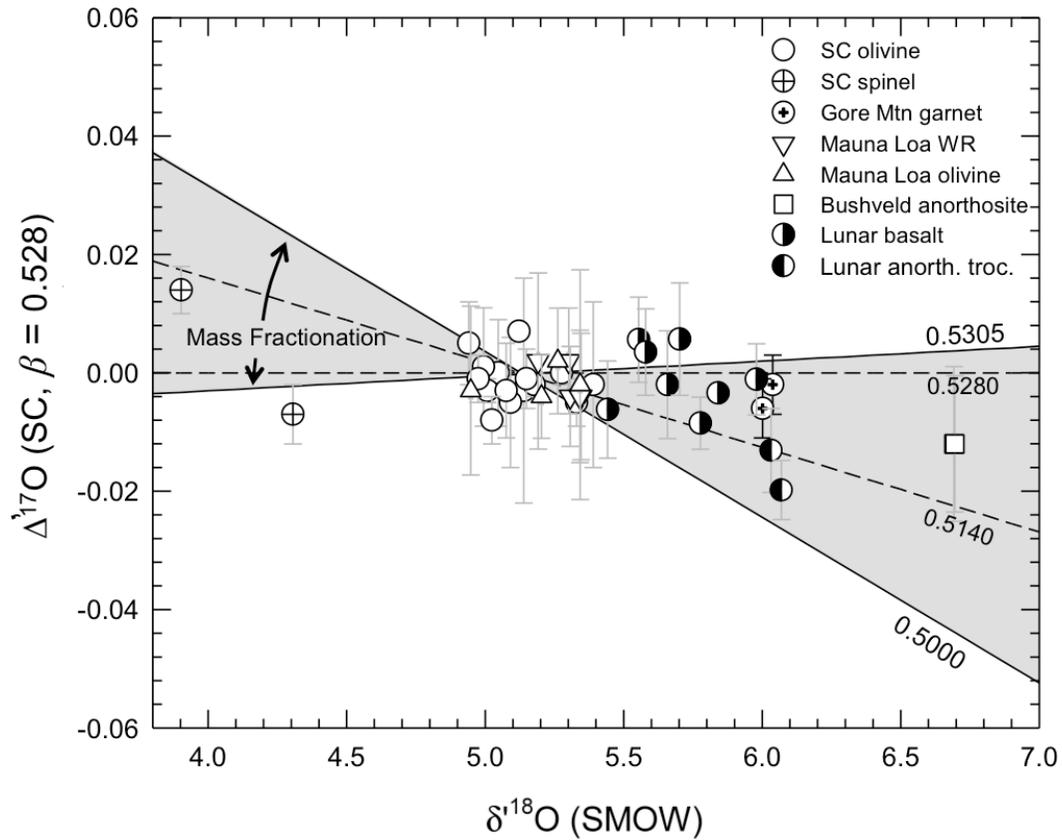

**Figure 2**. Plot of $\Delta'^{17}O$ vs. $\delta'^{18}O$ for lunar and terrestrial samples using a fractionation line with $\beta = 0.528$ passing through San Carlos olivine as the reference. Only the powders of lunar samples are plotted. The grey region indicates the regions accessible by mass fractionation starting from SC olivine. Different fractionation laws are labeled with their defining $\beta$ values. Error bars depict 2 standard errors for each measurement. Points lying inside of the grey region are consistent with simple one-stage mass-dependent isotope fractionation relative to SC olivine, implying they represent a single oxygen reservoir.

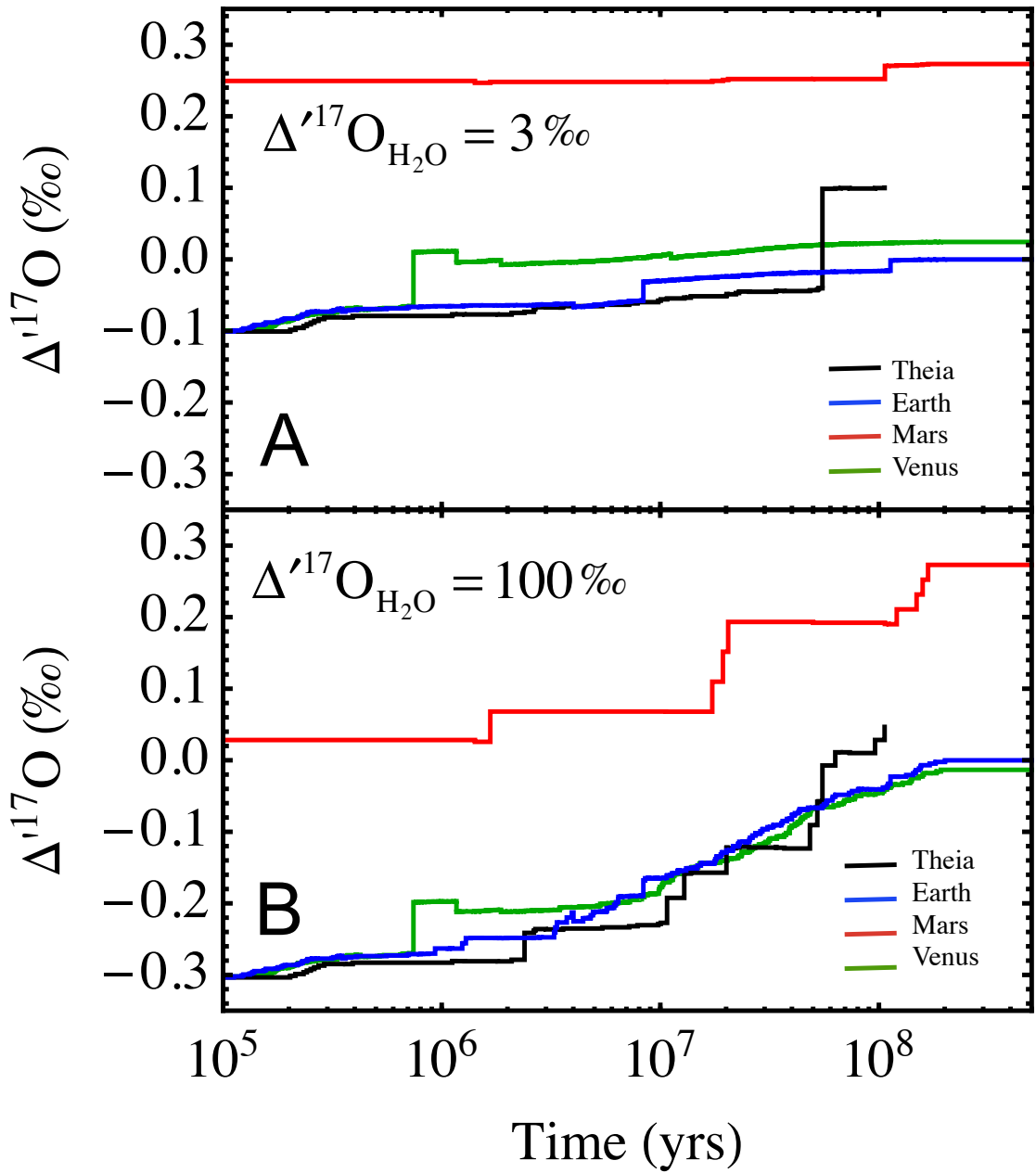

**Figure 3**. A simulation of the oxygen isotopic evolution of the terrestrial planets and last giant (Moon-forming) impactor, Theia. The $\Delta'^{17}O$ values of the growing Venus-like (green), Earth-like (blue), and Mars-like (red) planets are shown as a function of time as well as the value for the Theia-like impactor (black). (A) the case where the water oxygen reservoir has $\Delta'^{17}O = 3$ ‰. (B) the case where water $\Delta'^{17}O = 100$‰.